\documentclass[pmlr]{jmlr}
\usepackage{longtable}
\usepackage{booktabs}
\usepackage[load-configurations=version-1]{siunitx} % newer 

\theorembodyfont{\upshape}
\theoremheaderfont{\scshape}
\theorempostheader{:}
\theoremsep{\newline}

\jmlrvolume{106}
\jmlryear{2019}
\jmlrworkshop{Machine Learning for Healthcare}

 \usepackage{multirow}
\DeclareMathOperator{\E}{\mathbb{E}}

\title[Enhancing high-content imaging]{Enhancing high-content imaging for studying microtubule networks at large-scale}

\author{\Name{Hao-Chih Lee} \Email{hao-chih.lee@mssm.edu} \\
\Name{Sarah T. Cherng} \Email{sarah.cherng@mssm.edu} \\
\Name{Riccardo Miotto} \Email{riccardo.miotto@mssm.edu} \\
\Name{Joel T. Dudley} \Email{joel.dudley@mssm.edu} \\
\addr Institute for Next Generation Healthcare \\
Department of Genetics and Genomics Sciences\\
Icahn School of Medicine at Mount Sinai\\
New York, NY 10065, USA\\
} 
\begin{document}

\maketitle

\begin{abstract}
Given the crucial role of microtubules for cell survival, many researchers have found success using microtubule-targeting agents in the search for effective cancer therapeutics. Understanding microtubule responses to targeted interventions requires that the microtubule network within cells can be consistently observed across a large sample of images. However, fluorescence noise sources captured simultaneously with biological signals while using wide-field microscopes can obfuscate fine microtubule structures. Such requirements are particularly challenging for high-throughput imaging, where researchers must make decisions related to the trade-off between imaging quality and speed. Here, we propose a computational framework to enhance the quality of high-throughput imaging data to achieve fast speed and high quality simultaneously. Using CycleGAN, we learn an image model from low-throughput, high-resolution images to enhance features, such as microtubule networks in high-throughput low-resolution images. We show that CycleGAN is effective in identifying microtubules with 0.93+ AUC-ROC and that these results are robust to different kinds of image noise. We further apply CycleGAN to quantify the changes in microtubule density as a result of the application of drug compounds, and show that the quantified responses correspond well with known drug effects. 
\end{abstract}

\section{Introduction}
Microtubules, composed of tubulin, are the major part of the cytoskeleton and are crucial for cell division, migration, trafficking, and signalling \citep{lodish2008molecular}. In particular, microtubules play a central role in the process of cell mitosis, which has resulted in microtubule-binding agents (MTAs) acting as effective anticancer agents by inducing mitotic arrest and cell death as cancer cells divide more rapidly than normal cells \citep{mukhtar2014}. Continued study of microtubule responses to MTAs remain paramount due to significant discoveries related to difficulties in the use of MTAs to treat neoplastic cells, such as neuronal toxicity, and drug resistance \citep{orr2003}.  

High-throughput phenotypic screening is a powerful search tool for cancer drug candidates, which applies large-scale sets of compounds to cultured cells with the goal of quantifying their effects by measuring changes in cellular phenotypes. High-content screening (HCS) is a type of high-throughput phenotypic screening that uses microscopy to image cellular components and outputs images as measured phenotypes. This is often conducted with fluorescence microscopy, where cellular components are labeled with fluorescent probes and visualized under a microscope. HCS has been applied to identify potential drug candidates for treating cerebral cavernous malformation \citep{gibson2015strategy} and inhibiting bacterial growth \citep{sundaramurthy2013integration}.

Unbiased HCS requires the ability to capture a large number of images of subcellular outcomes within a reasonable amount of time. For practical reasons, trade-off decisions are made between imaging speed and imaging resolution for HCS-specific experiments \citep{buchser2014assay}. The consequences of such trade-offs are particularly challenging for HCS on microtubule networks for the purposes of drug discovery, as both image quality (i.e., recovering structures of microtubule networks) and speed (i.e., testing many compounds in sequence) are critical in the identification of effective drugs. The wide-field microscope, despite its superior speed, captures out-of-focus fluorescent signals that can greatly mask microtubules. Confocal microscopes are able to dramatically reduce fluorescent signals at the cost of slower image acquisition. 

\begin{figure}
  \centering 
  \includegraphics[width=\textwidth]{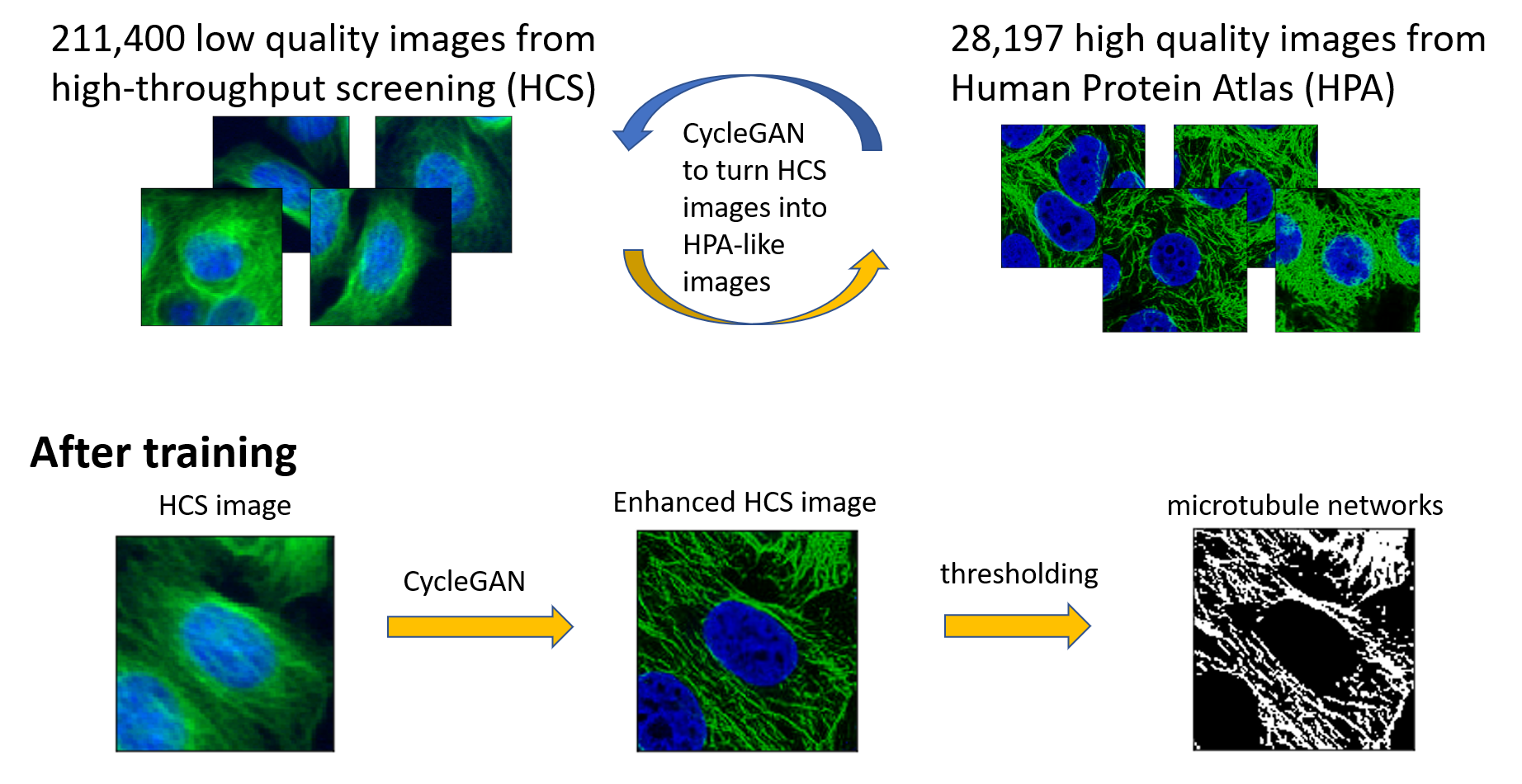} 
  \caption{Illustration of a CycleGAN and its application to enhance the quality of HCS images.}
  \label{fig:illustration} 
  \vspace{-4mm}
\end{figure}

In this paper, we propose a computational solution to capitalize on the advantages of both imaging modalities to generate high quality images at sufficiently fast speeds. To computationally enhance the important structural signals of microtubule networks within HCS images, we propose using a generative adversarial networks model approach (Figure \ref{fig:illustration}). We begin by training a model using high-resolution images of microtubule networks from low-throughput confocal images, collected from The Human Protein Atlas \citep{thul2017subcellular}. Using the CycleGAN framework \citep{zhu2017unpaired}, we demonstrate that the quality of HCS images can be enhanced by training a model to identify important features of microtubule networks from external reference images. With the removal of unwanted fluorescence-associated signals, the microtubule networks can be easily segmented from images using Otsu thresholding \citep{otsu1979threshold}. Finally, we apply this technique to verify whether expected drug effects on the density of microtubule networks can be observed from computationally enhanced HCS images.

\paragraph{Technical Significance}
The methodology presented here is an unsupervised approach to improve the quality of HCS images. The unsupervised nature of the proposed method is key to its applicability to HCS images, in which image noise may vary from batch to batch. The presence of such batch effects may lead to deteriorated performance of supervised models. We aim, specifically, to eliminate the need of supervised training using paired images, either taken simultaneously by another high-resolution imaging modality \citep{ouyang2018deep, zhang2019high, wang2019deep} or generated by {\it ad hoc} simulation \citep{nehme2018deep, weigert2018content}.

\paragraph{Clinical Relevance}
Our method offers new perspectives on the study of microtubule networks, a focal point of cancer drug discovery, by leveraging a fast but low-resolution imaging modality. This method can be generalized to other cellular components of interest with proper reference images.

\section{Related Work}
Various approaches have been developed to improve the quality and resolution of fluorescent microscopy. A super-resolution microscopy is the current state-of-the-art method, which breaks down the diffraction barrier using a specialized hardware design. For instance, simulation emission depletion microscopy uses a structured laser light to selectively deactivate fluorophores around a focal point to improve resolution \citep{hell1994breaking}. Photoactivated localization microscopy (PALM, \cite{hess2006ultra}) and stochastic optical reconstruction microscopy (STORM, \cite{bates2007multicolor}) record time-lapsed images of blinking fluorophores to computationally localize proteins with nanometer accuracy. However, super-resolution microscopy is not a viable modality for high-throughput drug screening due to speed constraints. 

Enhancing the resolution of natural images is long rooted in the field of computer vision \citep{yang2014single, nasrollahi2014super}. Recently, neural network models have been applied to enhance the resolution of natural photography \citep{dong2014learning, ledig2017photo}. This method was subsequently applied to fluorescent microscopy. For example, a supervised neural network can be trained to transform low resolution images into high resolution ones, either taken simultaneously by a different imaging modality, or generated by simulation. The content-aware image restoration (CARE) networks were trained on pairs of registered images of both high and low signal-to-noise ratios (SNR) to enable restoring biological structures from low SNR images \citep{weigert2018content}. Additionally, \citeauthor{wang2019deep} trained a convolutional neural network using images taken under different imaging modalities to improve image resolution. Deep STORM \citep{nehme2018deep} and ANNA-PALM \citep{ouyang2018deep} both utilized paired low-resolution images of blinking emitters and their coordinates to increase speed for the reconstruction of super-resolved images.

Three strategies are generally used to quantify the structure of microtubules from images. The first approach utilizes direct measurements from image segmentation of microtubules. For example, \citeauthor{faulkner2017automated} applied local thresholding to segment microtubules and analyze the length of its morphological skeleton. \citeauthor{jacques2013microfilament} developed a line search algorithm to connect neighboring pixels with comparable values to detect microtubules. This approach relies on the successful identification of individual microtubules, which can be challenging for HCS imaging due to its susceptibility to low image quality \citep{boudaoud2014fibriltool}. The second approach attempts to quantify the local coherent pattern of microtubules via image textures, utilizing patterns extracted from image gradients \citep{karlon1999measurement},  nematic tensors \citep{boudaoud2014fibriltool, tsugawa2016extracting} and Haralick textures \citep{berges2017toward}. The third approach is the generative modeling approach that searches for parameters to simulate the most realistic images of microtubule networks \citep{shariff2010generative}. Our work aims at improving the applicability of the first approach by enhancing the quality of HCS images.

\section{Methods}
\subsection{CycleGAN for enhancing fluorescent images}
Generative adversarial networks (GANs, \cite{goodfellow2014generative}) are a class of neural network models that map samples from a prior distribution to the data distribution. GANs consist of two neural networks: a discriminator $D(x)$ and a generator $G(z)$. The generator performs the primary task of mapping a sample from the prior distribution $p_z$ to a data point while the discriminator classifies how similar this generated data point is compared to the real data. Both neural networks are trained simultaneously by solving a min-max game, in which the generator seeks to create samples with with the goal of fooling the discriminator. The discriminator seeks to learn features of the generated data so that it is able to discriminate generated samples from real data points. The objective function is originally formulated as
\[\mathcal{L}_{GAN}(D, G) = \E_{x\sim data }\log(D(x)) +  \E_{z \sim p_z }\log(1-D(G(z))).\] It has been shown that GANs can generate very realistic images, with greater clarity than ones generated by autoencoders \citep{goodfellow2014generative}. \\ 

While GANs are designed as a random sampler from the data distribution, conditional GANs generate images based on input images \citep{isola2017image}. The generator of the conditional GAN learns a mapping from an input image $x$ and a random noise vector $z$, to a generated image $y$, while the discriminator remains the same as in the original GAN. To enforce the consistency between generated images $y$ and input images $x$, \citeauthor{isola2017image} proposes the use of the $L_1$ discrepancy to penalize the model, i.e., \[\mathcal{L}_{L_1}(G) = \E_{x,y,z}\|G(x,z) - y\|_1,\]
which requires supervised pairs of trained images $x$ and $y$. The conditional GAN is then trained to minimize the objective function
\[\mathcal{L}_{cGAN}(D, G) = \mathcal{L}_{GAN}(D, G) + \lambda \mathcal{L}_{L_1}(G)\]
In the context of image-to-image translation, this model is often referred to as the ``pix2pix" model.

CycleGANs take conditional GANs one step further. The primary goal of a cycleGAN is to transform images from a source distribution to an image from a target distribution without paired training examples \citep{zhu2017unpaired}. For example, CycleGAN, trained on paintings of an artist, can render natural photographs into the style of a given artist. CycleGAN consists of four neural networks: two generators $G_{A\to B}(x)$ and $G_{B\to A}$(x) that map the source and target distributions from one to another and two discriminators $D_A(x)$ and $D_B(x)$ to judge how realistic the generated samples are. To ensure that the generators retain information of its domain images, \citeauthor{zhu2017unpaired} proposed using cycle-consistency loss:
\[\begin{array}{ll}\mathcal{L}_{\textup{consistency}} = &\E_{A\sim \textup{source images }}\|G_{B\to A}(G_{A\to B}(A)) - A\|_1\\
& + \E_{B \sim \textup{target images} }\|G_{A\to B}(G_{B\to A}(B)) - B\|_1,\end{array} \]
which measures only self-reconstruction errors and thus circumvents the need of supervised paired training samples. The CycleGAN is trained to minimize the objective function
\[\mathcal{L}_{obj} = \mathcal{L}_{GAN}(D_A, G_{B\to A}) + \mathcal{L}_{GAN}(D_B, G_{A\to B}) + \lambda \mathcal{L}_{\textup{consistency.}}\]

One of the fundamental assumptions of CycleGANs is that there exists a relationship between the source and target domains. For example, two pictures with different styles are rendered differently, despite capturing the same object. Such an assumption would hold true for images of the same cell samples taken under different microscopes. Images from a microscope can be seen as the output of an image formation process which takes labeled biological structures as input and returns an array of pixel intensity. The image formation processes of two different imaging modalities (e.g., HCS and high-resolution confocal images) $A$ and $B$ can be modeled as $f_A(X)$ and $f_B(X)$ respectively, where $X$ indicates a potential configuration of the biological structures of interest. Assuming we can reverse the image formation process $f_B$ by a reconstruction algorithm, taking the composition of $f_A\circ f_B^{-1}$ could, in principle, transfer a image taken under one imaging modality $B$ to an image taken under a different imaging modality $A$ and vice-versa. The generators of the CycleGAN can thus approximate the composition functions $G_{A\to B}\approx f_B\circ f_A^{-1}$ and $G_{B\to A}\approx f_A\circ f_B^{-1}$ without solving the reconstruction inverse problems. Though this theoretically justifies the use of CycleGAN, we seek to design a prototype and evaluate its empirical performance on enhancing fluorescent imaging.

\subsection{Data Collection}
\begin{figure}[t]
  \centering 
  \includegraphics[width=\textwidth]{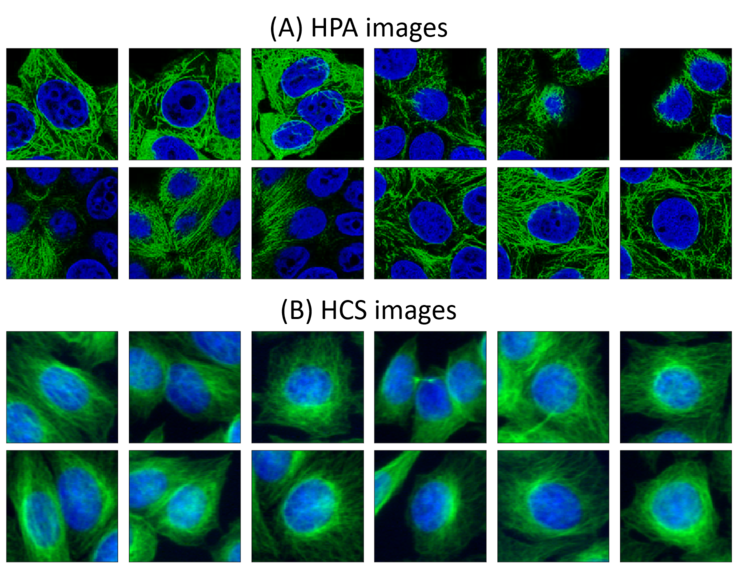} 
  \caption{Examples of the (A) human protein atlas (HPA) and (B) high-content screening (HCS) image patches. Cell nuclei and microtubules are shown in blue and green, respectively.}
  \label{fig:HAP_HCS_example} 
  \vspace{-4mm}
\end{figure}

We use images from The Human Protein Atlas to curate a reference set of high resolution images. The Human Protein Atlas is a database that archives images of 12,073 protein taken from 64 cell lines \citep{thul2017subcellular}. Among these images, the cell nucleus and microtubules are typically imaged at the same time as the reference coordinate for other proteins of interest. Images are collected under a $63\times$ confocal microscope, which reduces out-of-focus light. We select 1,836 images taken from the MCF7 cell line, which is the same cell line used in our HCS image data. Images are down-sampled by 4-fold and re-sized from $2048\times 2048$ to $512\times 512$, from which we generate single-cell image patches. We first identify the cell nucleus using Ostu segmentation \citep{otsu1979threshold}. Images patches of size $128\times 128$, centered at the centroid of the cell nucleus, are then cropped. We only keep image patches that fall entirely within the original image. In total, we collect 28,197 single cell image patches. Each image patch contains a nucleus channel and a microtubule channel. Figure \ref{fig:HAP_HCS_example} shows examples of the high resolution image patches described here. We refer to this set of high-resolution image patches as the ``HPA images" moving forward.

We also process the image set BBBC021v1 \citep{caie2010high}, available from the Broad Bioimage Benchmark Collection \citep{ljosa2012annotated}. BBBC021v1 is a HCS experiment on MCF7 cells. 13,200 images are collected using a 20x wide-field microscope. Each image contains a nucleus, microtubule, and actin filament channel. We retain only the nucleus and microtubule channels to match the settings of our HPA image set. Using the same procedure to select images for the HPA set, we generate 84$\times$84 images patches centered at the cell nucleus from untreated cells. In total, we collect 211,400 single-cell image patches from cells without the application of drug treatment for model training. We choose the size of image patches so that physical size of cell nucleus matches with those in our HPA images, as explained in the next section. We refer to this set of high-throughput, wide-field microscope images, as the ``HCS images" set.

\subsection{Implementation details}
\subsubsection{Data augmentation}

For both training and inference, we first re-size the low-resolution HCS images from $84\times84$ to $128\times 128$ to ensure that the average size of cell nuclei in the HPA and HCS image sets are roughly equivalent. The magnification factor is set to $128/84\approx 1.523$, which we determine by comparing the average size of cell nuclei in both HPA and HCS images. We apply Otsu thresholding to both image sets and determine that, on average, a nucleus contains 2,172 and 957 pixels for HPA and HCS images, respectively and therefore $128 \times \sqrt{957/2172} \approx 84$. We then randomly crop $128\times 128$ images into $64\times 64$ sized images when training.

\subsubsection{Parameters for training}
We use a residual network with 9 repeated blocks for the generator and a 4-layer fully-connected convolutional neural network, sharing the same architecture as the 70$\times$70 PatchGAN for the discriminator \citep{isola2017image, zhu2017unpaired}. The regularization parameter $\lambda$ is set to 10. We apply the ADAM optimizer \citep{kingma2014adam} for all experiments. We use a constant step size of 0.0002 for the first 20 epochs and then linearly reduce step size to 0 for the next 20 epochs. A batch of 8 samples is used for each iteration. 

\section{Results and Discussion}
\subsection{Evaluation by simulation}
We conduct 3 simulations to evaluate the performance of the CycleGAN model. The source code is available at \url{https://github.com/DudleyLab/widefield2confocal}. 
\begin{enumerate}
\item {\bf Case 1 (blurred images)}: We randomly split HPA images into 45/45/10 training, validation and testing sets. The training set contains 12,644 unprocessed, high-quality images while the validation and testing set contain 12,661 and 2892 degraded images, respectively. Using training and validation sets for model training, our goal is to evaluate the quality of reconstructed images from the testing set. The train/validate/test split is conducted on raw images that are downloaded from the Human Protein Atlas, but not cropped image patches, to avoid overlapping regions that may occur in the three sets of images. The 12,661 image patches in the validation set and 2,892 image patches in the testing set are down-sampled from 128$\times$128 to 84$\times$84 image patches, and blurred with a Gaussian kernel (bandwidth = 1), with the addition of Gaussian noise (standard deviation = 3). Specifically, we simulate a degraded image $I_{simulated}$ by
\[I_{simulated} = g_{\sigma=1} * I_{tubule} + 3 \cdot \epsilon \]
where $g_{\sigma=1}$ is the Gaussian kernel with a bandwidth of 1, $\epsilon$ is the normal Gaussian noise, and $*$ is the convolution.

\item {\bf Case 2 (bleed-through signals)}: We keep the same train/validate/test split as in case 1. In addition to down-sampling, blurring, and the addition of noise, we introduce a bleed-through signal to the microtubule channel from the nucleus channel \citep{buchser2014assay}. 
Specifically, we simulate an image $I_{simulated}$ by
\[I_{simulated} = g_{\sigma=1} * I_{tubule} + p\cdot(g_{\sigma=5} * I_{nucleus}) + 3 \cdot \epsilon \]
where $g_{\sigma=1}$ and $g_{\sigma=5}$ are Gaussian kernels with bandwidths of 1 and 5, respectively, and $\epsilon$ is normal Gaussian noise. We use $p=0.2,$ 0.5, and a random number ranging from 0.2 to 0.5 for cases 2a, 2b, and 2c respectively.  

\item {\bf Case 3 (diffusive fluorescent probes)}: We simulate the impact of non-specific signal from diffusive dyes \citep{toseland2013fluorescent}. The simulated images are generated as 
\[I_{simulated} = (1-p)(g_{\sigma=1} * I_{tubule}) + p\cdot(g_{\sigma=5} * I_{tubule}) + 3 \cdot \epsilon.\]
The first term is the true microtubule signal convoluted with the point spread function. The second term simulates the non-specific signal from dyes diffusing away from the microtubules. We simulate this effect by convolving true microtubule signals with a larger Gaussian kernel. We use the weighted average of two simulated signals to ensure that the pixel values remain in the same range. We use $p=0.2,$ 0.5, and a random number ranging from 0.2 to 0.5 for cases 3a, 3b, and 3c respectively. 
\end{enumerate}
All simulated images are binned as 8-bit integers.

\begin{figure}[t]
  \centering 
  \includegraphics[width=\textwidth]{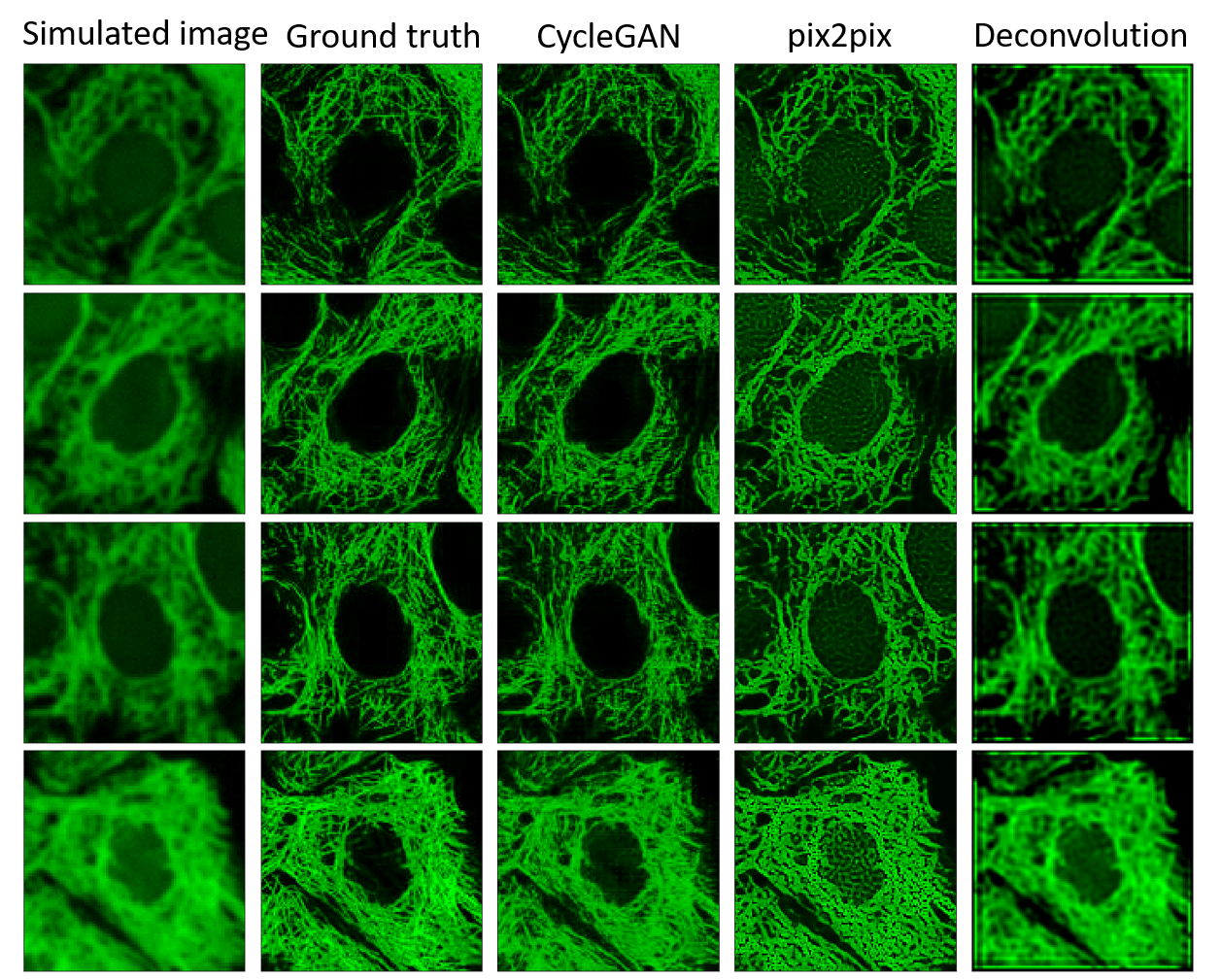}
  \caption{Examples of reconstructed images in case 2c.}
  \label{fig:sim_rec_2c} 
  \vspace{-4mm}
\end{figure} 

\begin{figure}[t]
  \centering 
  \includegraphics[width=\textwidth]{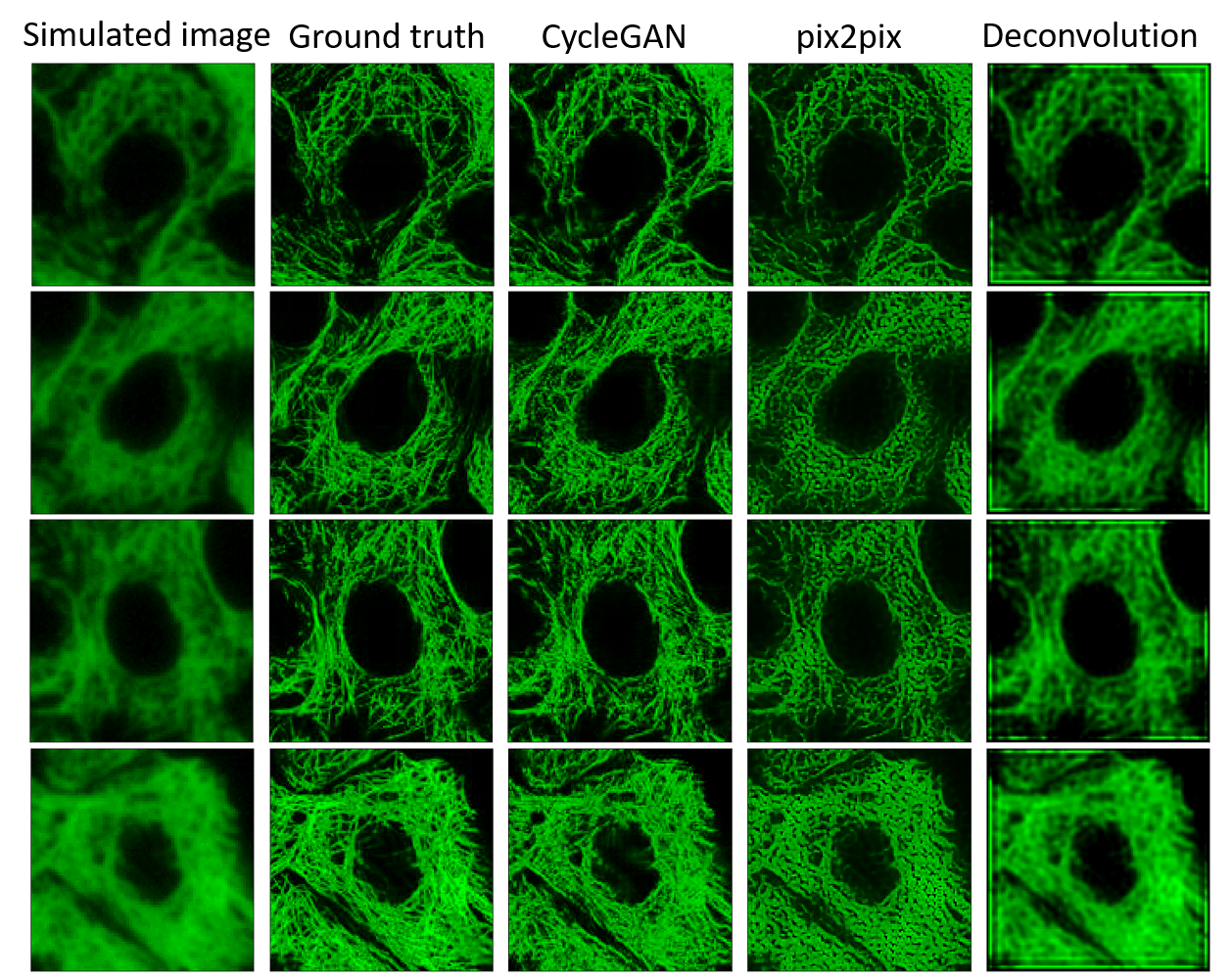}
  \caption{Examples of reconstructed images in case 3c.}
  \label{fig:sim_rec_3c} 
  \vspace{-4mm}
\end{figure}

Using the training and validation sets, we train a CycleGAN model for each simulation separately. After training, each model is applied to the corresponding testing images to recover microtubule signals. For comparison, we use the pix2pix model and the Richardson-Lucy deconvolution algorithm as baseline methods. For the pix2pix model, we create a set of simulated images from training images using the same setting as in case 1. A pix2pix model is then trained using only the training images and the corresponding simulation images. This pix2pix model is then applied to other testing image sets to recover microtubule signals. We also apply the Richardson-Lucy deconvolution algorithm to test images using $g_{\sigma = 1}$ as the point spread function. For Richardson-Lucy deconvolution, we use scikit-image implementation (\cite{van2014scikit}). Figure \ref{fig:sim_rec_2c} and \ref{fig:sim_rec_3c} present examples of simulated images and reconstructed images using methods that we describe in case 2c and 3c.

We use three metrics to measure performance, with original test images acting as ground-truth.
\begin{enumerate}
\item Structural similarity index (SSIM, \cite{wang2004image}) and the multi-scale structural similarity index (MSSSIM, \cite{wang2003multiscale}). SSIM and MSSSIM are scores designed to measure the perception-based similarity of two images in terms of luminance, contrast, and structure, but not the absolute difference in pixel values. Both scores are commonly used to benchmark the quality of images generated by GANs \citep{kurach2018gan}. SSIM and MSSSIM both range from -1 and 1. A higher score indicates greater similarity between two images.

\item We also evaluate whether the intensity of reconstructed signals indicate the true existence of microtubules. We label pixels in an image with values larger than the average as positive. The pixel value of the reconstructed image is then used as a logit to predict positive labels. Finally, the area under ROC-curve (AUC-ROC) is computed as a metric to measure prediction accuracy.

\end{enumerate}
The results our case studies are shown in table \ref{tab:performance}. In the first case (case 1), the pix2pix model outperforms the other two. However, in the presence of bleed-through signals  (case 2) and diffusive fluorescent probes (case 3), the CycleGAN performs best. This suggests that the pixel2pixel model and Richardson-Lucy deconvolution, which both assume that blurring is due only to noise, may fail to handle noise sources that deviate from the designed image formation model (Figure \ref{fig:sim_rec_2c} and Figure \ref{fig:sim_rec_3c}). On the contrary, CycleGAN can robustly adapt to three types of noise sources. In all three cases, the 0.93+ AUC-ROC achieved by our CycleGAN model suggests that recovered pixel values are informative to the presence of microtubules.

% \usepackage{multirow}
\begin{table}[ht]
\centering 
\begin{tabular}{|c|c|c|c|c|}
\hline
                              &         & CycleGAN                            & pix2pix                            & deconvolution              \\
\hline\hline\multirow{3}{*}{\begin{tabular}{c} Case 1 \\ blurred images \end{tabular}}  & SSIM    & $65.8\pm5.4$           & {$\bf 70.6\pm5.6$} & $65.5\pm4.9$  \\
                              & MSSSIM  & $74.9\pm4.5$ & { $\bf 77.1\pm4.4$} & $72.0\pm4.8$  \\
                              & AUC-ROC & $94.1\pm2.2$           & { $\bf 94.8\pm1.9$} & $93.6\pm2.1$  \\
\hline
\multirow{3}{*}{\begin{tabular}{c} Case 2a \\ bleed-through\\  signals, p= 0.2 \end{tabular}} & SSIM    & {$\bf 62.2\pm5.3$}  & $52.5\pm7.2$          & $50.8\pm5.9$  \\
                              & MSSSIM  & $\bf 73.8\pm4.5$ & $64.5\pm6.3$ & $65.2\pm4.7$  \\
                              & AUC-ROC & {$\bf 93.6\pm4.8$}  & $89.6\pm6.6$          & $90.6\pm5.9$  \\
\hline
\multirow{3}{*}{\begin{tabular}{c} Case 2b \\ bleed-through\\  signals, p = 0.5 \end{tabular}} & SSIM    &  {$\bf 61.4\pm4.8$}  & $47.0\pm8.7$          & $47.0\pm6.8$  \\
                              & MSSSIM  & $\bf 73.3\pm4.0$ & $57.5\pm8.7$ & $60.9\pm6.1$   \\
                              & AUC-ROC & {$\bf 93.5\pm3.6$}  & $83.8\pm9.1$          & $84.8\pm9.2$  \\
\hline
\multirow{3}{*}{\begin{tabular}{c} Case 2c \\ bleed-through\\  signals, random p \end{tabular}} & SSIM    & {$\bf 61.6\pm5.5$}  & $50.0\pm8.3$          & $49.2\pm6.5$  \\
                              & MSSSIM  & $\bf 73.2\pm4.9$ & $61.3\pm8.1$ & $63.5\pm5.5$ \\
                              & AUC-ROC & {$\bf 93.4\pm5.0$}  & $87.1\pm8.3$          & $88.4\pm7.8$  \\
\hline\multirow{3}{*}{\begin{tabular}{c} Case 3a \\ diffusive fluorescent \\  probes, p = 0.2 \end{tabular}} & SSIM    & {$\bf 63.5\pm5.7$}  & $61.1\pm6.2$          & $55.3\pm5.3$  \\
                              & MSSSIM  & $\bf 73.0\pm4.9$ & $71.5\pm5.1$ & $64.8\pm5.4$  \\
                              & AUC-ROC & {$\bf 93.8\pm2.3$}  & $93.2\pm2.3$          & $92.6\pm2.3$  \\
\hline\multirow{3}{*}{\begin{tabular}{c} Case 3b \\ diffusive fluorescent \\  probes, p = 0.5 \end{tabular}} & SSIM    & {$\bf 59.7\pm5.8$}  & $46.5\pm7.8$          & $43.4\pm6.2$  \\
                              & MSSSIM  & $\bf 70.0\pm5.1$ & $59.1\pm6.5$ & $54.6\pm6.4$  \\
                              & AUC-ROC & {$\bf 93.0\pm2.7$}  & $90.0\pm3.2$          & $91.1\pm2.7$  \\
\hline\multirow{3}{*}{\begin{tabular}{c} Case 3c \\ diffusive fluorescent \\  probes, random p \end{tabular}} & SSIM    & {$\bf 61.0\pm7.2$}  & $55.4\pm8.6$          & $50.6\pm7.1$  \\
                              & MSSSIM  & $\bf 71.2\pm5.9$ & $66.9\pm7.1$ & $60.9\pm6.8$  \\
                              & AUC-ROC & {$\bf 93.3\pm2.6$}  & $92.0\pm2.9$          & $92.1\pm2.5$ \\
\hline
\end{tabular}
 \caption{Performance of image reconstruction.}
 \label{tab:performance} 
 \vspace{-4mm}
\end{table}

Figure \ref{fig:performance_over_iter} presents the performance of our CycleGAN during training iterations in cases 2 and 3. In all simulated scenarios, CycleGAN achieves high accuracy in the first 4 epochs. Accuracy then either plateaus or improves gradually moving forward. We do not not observe over-fitting within 40 epochs.

\subsection{Application to high-throughput screening}
\begin{figure}
  \centering 
\begin{minipage}[c]{\textwidth}
\centering
   \includegraphics[width=\textwidth]{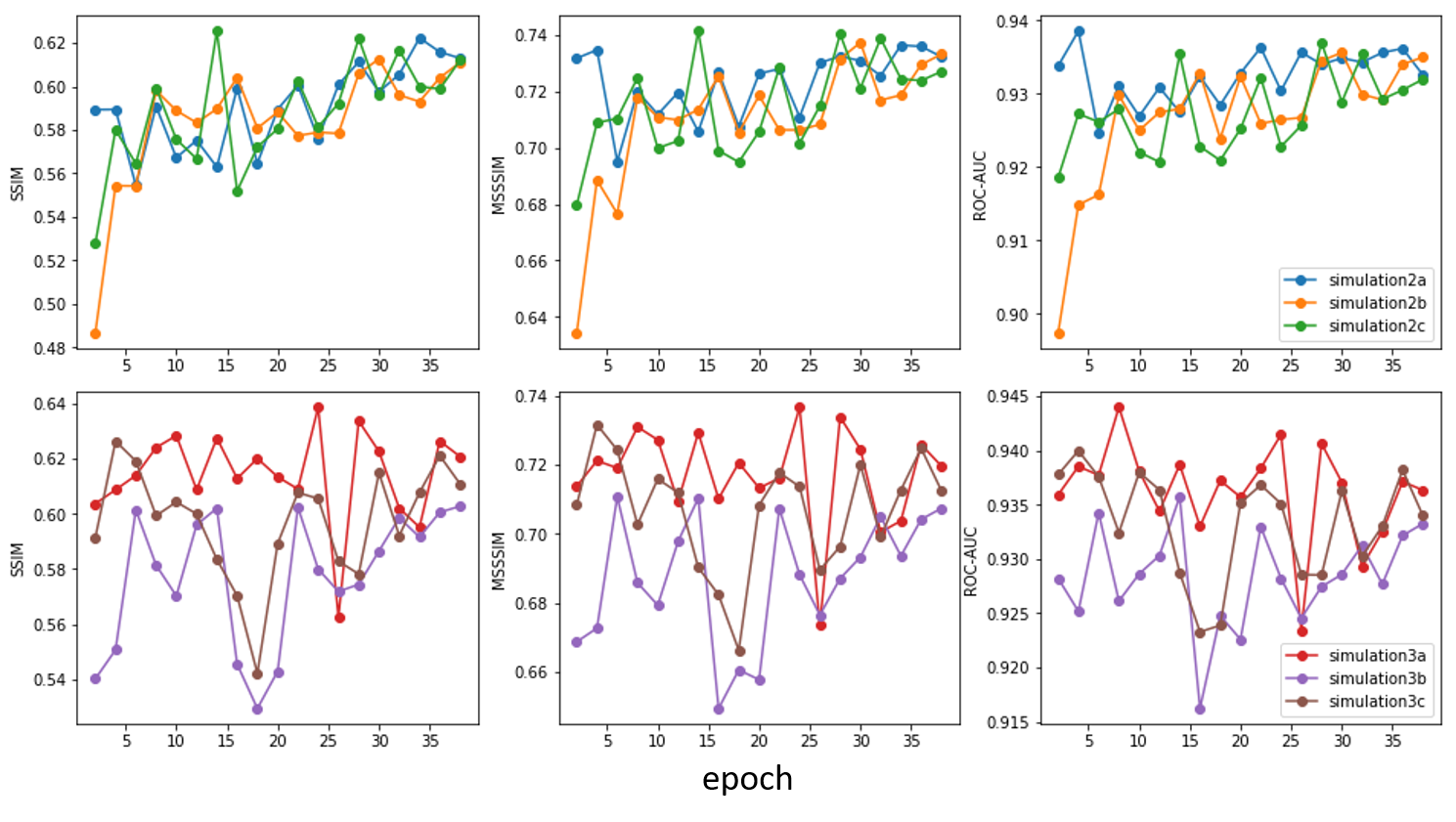}
    \caption{Performance during training iterations.}
    \label{fig:performance_over_iter} 
\end{minipage}
\begin{minipage}[c]{\textwidth}
\centering
  \includegraphics[width=\textwidth]{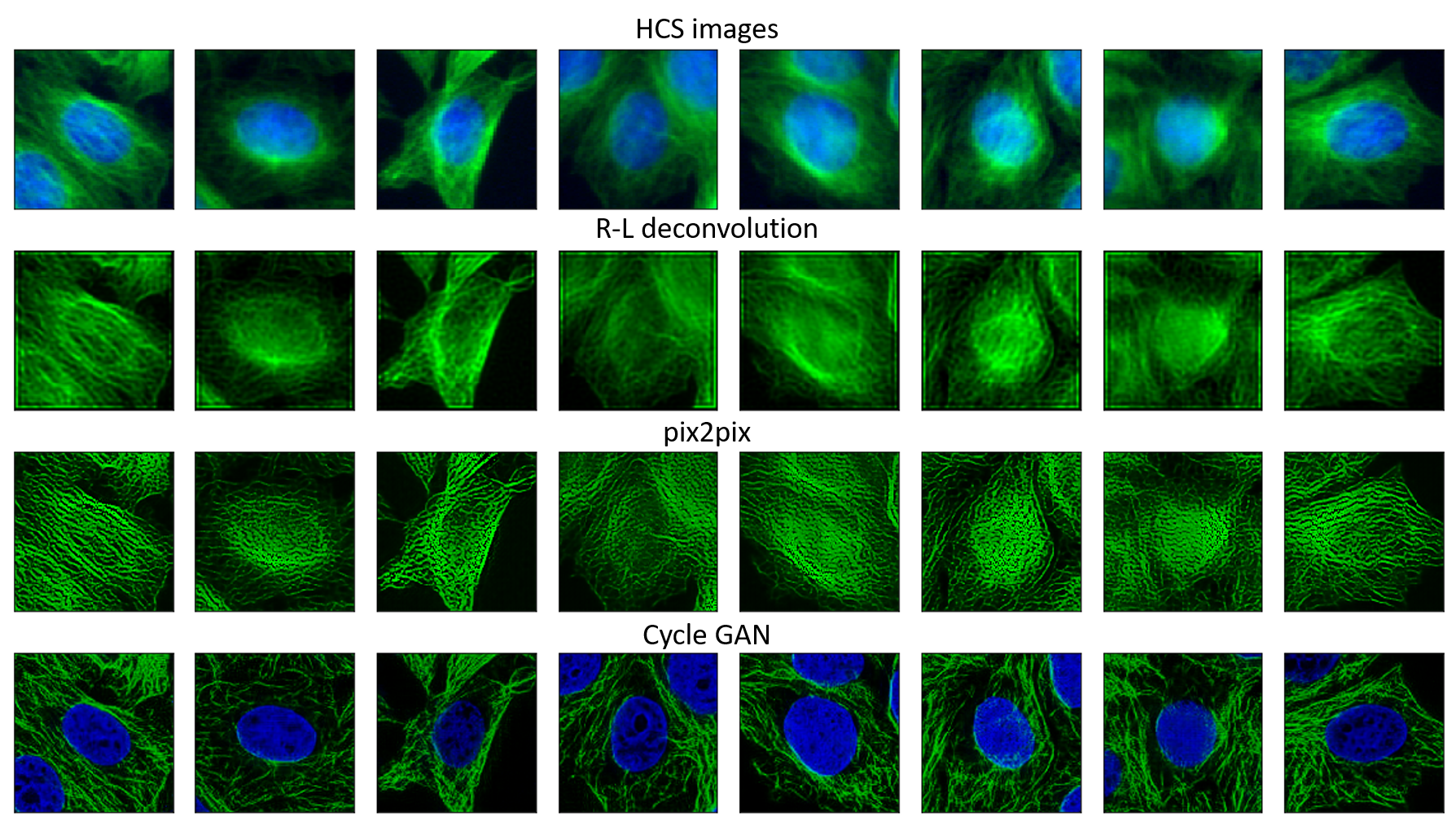} 
  \caption{Examples of reconstructed HCS images. The Richardson-Lucy deconvolution and pix2pix model are applied only to the microtubule channel while the CycleGAN is trained using both the microtubule and nucleus channel. Cell nuclei and microtubules are shown in blue and green respectively.}
  \label{fig:tub_rec} 
  \vspace{-2mm}
\end{minipage}

\end{figure}

Next, we apply our CycleGAN model to HCS images. We train a CycleGAN using 211,400 HCS image patches of untreated cells as the source images, and 28,197 HPA image patches as the target images. Figure \ref{fig:tub_rec} presents examples of enhanced images. We note two qualitative observations:

\begin{enumerate}
\item CycleGAN seems to adjust uneven illumination. Microtubule networks are generally denser around the cell nucleus than on the periphery. As a result, the fluorescent signals are generally weaker on the periphery in HCS images. CycleGAN is able to adjust this uneven illumination.
\item However, CycleGAN is also prone to eliminate microtuble signals within the region of a cell nucleus. This seems to be a biased correction resulting from using HPA images as a prior model since microtubules are also rarely present within the region of a cell nucleus in HPA images.
\end{enumerate}

\begin{figure}
\begin{minipage}[c]{\textwidth}
  \centering 
  \includegraphics[width=\textwidth]{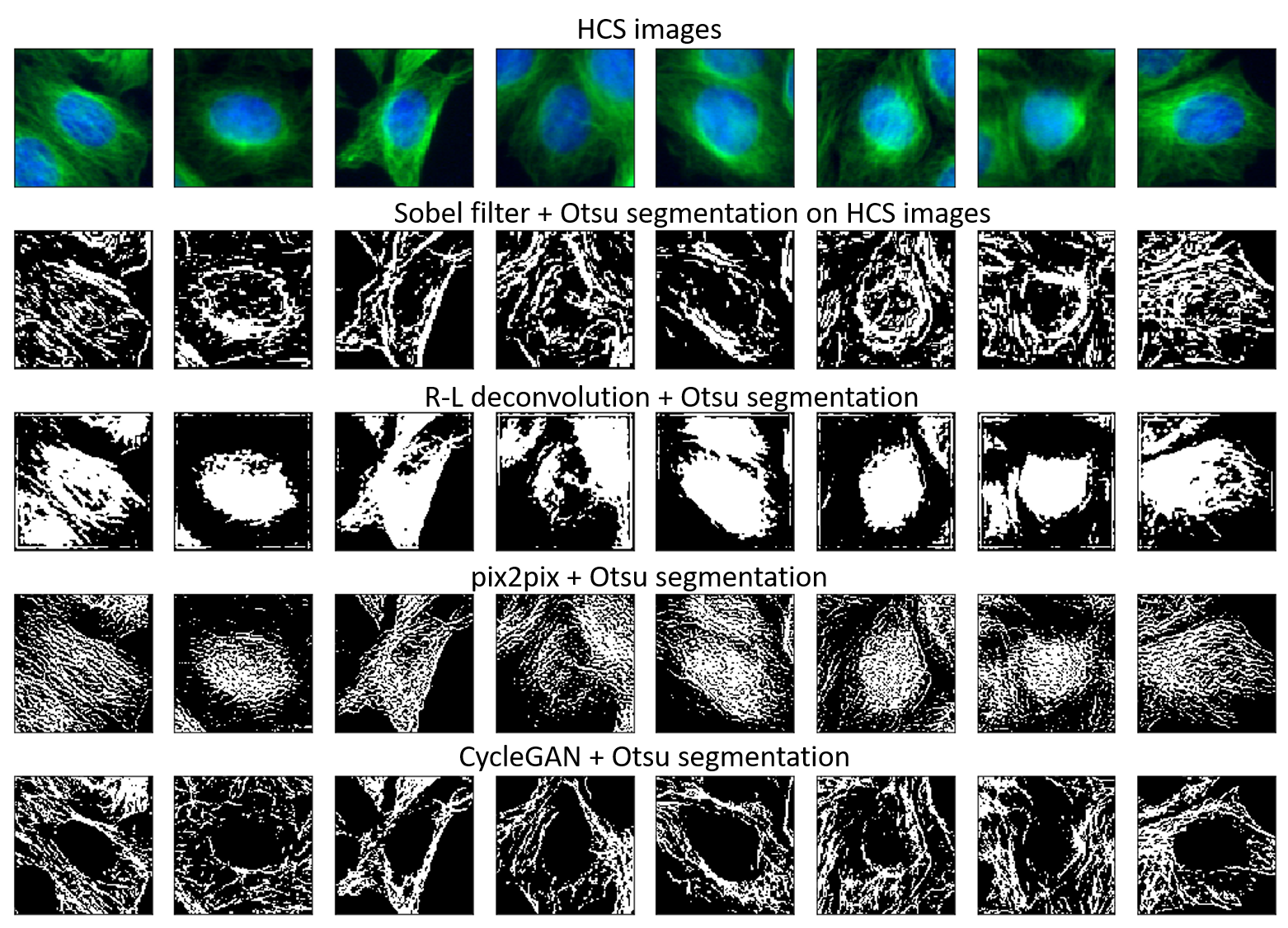} 
  \caption{Examples of segmented microtubule networks.}
  \label{fig:tub_seg} 
  
\end{minipage}
\vspace{2mm}
\begin{minipage}[c]{\textwidth}
  \centering 
  \includegraphics[width=\textwidth]{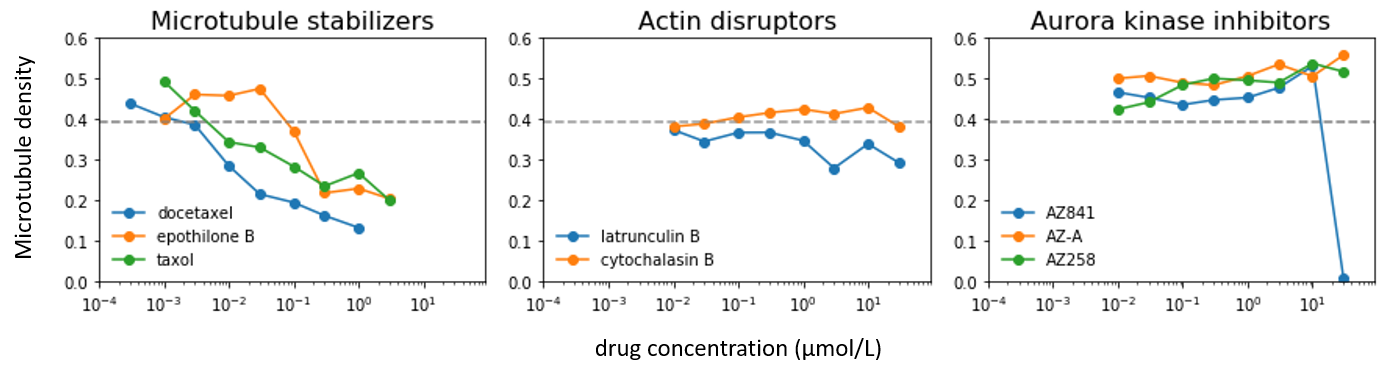} 
  \caption{The dose-response curves of the effects of 8 drugs on microtubule density.}
  \label{fig:drug_eff} 
\end{minipage}

\end{figure}

Based on the first observation of CycleGAN correcting uneven illumination, it is possible to segment microtubules from enhanced images using Otsu thresholding. Figure \ref{fig:tub_seg} shows examples of segmented microtubule networks. As a baseline, we apply the Sobel edge detector, followed by Otsu thresholding, to the original HCS images. Images that are reconstructed by both Richardson-Lucy deconvolution and pix2pix models, which are both trained using the first simulation case, are also segmented by Otsu thresholding for comparison. While there are practical difficulties in quantifying the accuracy of segmenting microtubule networks in HCS images, we note that the AUC-ROC metric described in section 4.1 evaluates the performance of this task using simulated data.

Lastly, we demonstrate an application of the proposed method to quantify drug effects on microtubule networks. 
We first apply CycleGAN to images of drug-treated cells and obtain segmented microtubule networks using Otsu segmentation. We then computed the number of segmented pixels that are no more than 8 pixels away from nuclei. The microtubule density is then defined as the ratio of the number of segmented pixels to the total area of this perinuclear region, excluding the region of cell nucleus itself. We restrict measurement within the perinuclear regions to control the cell size as a variable that can confound the estimation of microtubule density. 200 cells are randomly selected for each drug-concentration combination to estimate the drug effect on microtubule density.

Figure \ref{fig:drug_eff} shows the average microtubule density from cells treated with drugs of 3 known mechanisms of action. The dashed line represents the average microtubule density computed from 1,000 untreated cells. 
We observe that microtubule stabilizers, which are chemicals that interfere the growth and shrinkage of microtubules \citep{yvon1999taxol, jordan2004microtubules} and promote microtubule polymerization \citep{jordan1993mechanism}, induce higher microtubule density at low dosages, followed by a steep decline at high dosages. We note that the sharp drop to low microtubule density at high drug concentration levels is due to the increasing number of dying cells with shrunken cytoplasms (supplementary figure 1). We observe minor effects on microtubules for actin disruptors (supplementary figure 2). Lastly, we observe elevated microtubule density on cells treated with the Aurora kinase inhibitor, as Aurora kinase inhibitors seem to interrupt mitotic progression that leads to multinucleation (supplementary figure 3).

\section{Conclusions and Future Work} 
In this paper, we present a framework to enhance the quality of HCS images using CycleGAN. Our empirical results demonstrate that CycleGAN can learn an unsupervised image model from high-quality images to identify relevant signals in low-quality images. In particular, we also show that CycleGAN can robustly adjust different types of image noise to achieve accurate identification of microtubules. Despite its superior performance, we observe biased removal of microtubules within the region of the cell nucleus, a bias likely inherited from curated reference images that we plan to resolve in future work.

In applying our method to HCS data, it is possible to identify hundreds of thousands of segmented microtubule networks from drug-treated cells. Here, we specify a case study to demonstrate the potential of this method by first analyzing the microtubule density. We plan to incorporate feature extraction techniques such as Hough transformation, to further analyze the lengths, orientations, and numbers of microtubules under different drug treatments. This new resource will allow us to discover new insights into the essential role of microtubule networks in cancer drug discovery research. Our method offers an opportunity for large-scale mining of microtubule networks. Importantly, this method is not limited to the study of microtubules alone. Indeed, with appropriate low-throughput, high resolution training images, the methods presented here may also effectively sharpen images of other subcellular components of interest. 
% ACKNOWLEDGEMENTS ONLY GO IN THE CAMERA-READY, NOT THE SUBMISSION
\acks{The authors would like to thank support from the Hasso Plattner Foundation and a courtesy GPU donation from NVIDIA.}

\bibliography{main}

\newpage
\appendix
\section*{Appendix}
\setcounter{figure}{0}   

\begin{figure}[!htb]
  \centering 
  \includegraphics[width=\textwidth]{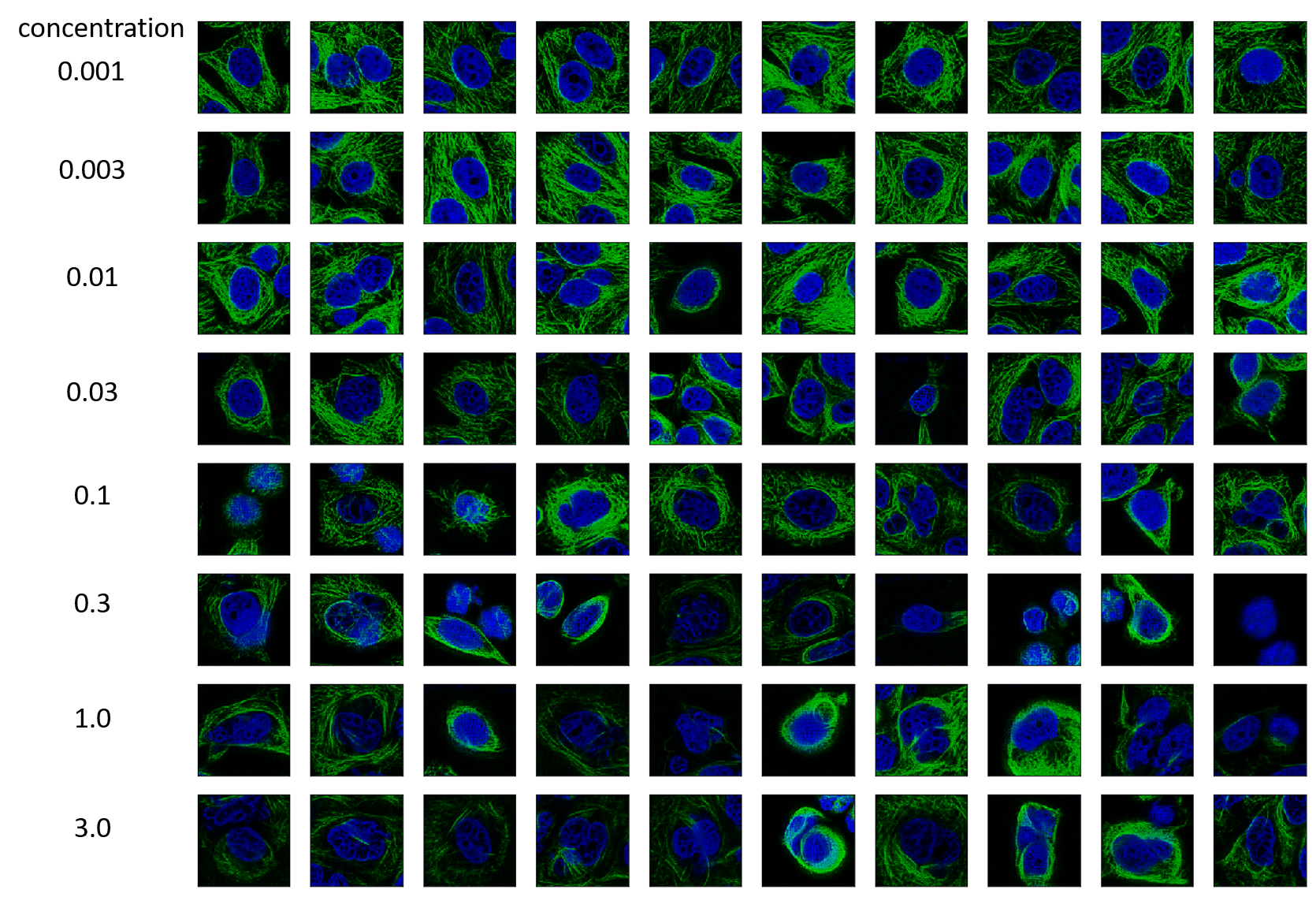} 
  \caption{Examples of reconstructed images from cells treated with Epothilone B.}
  \label{fig:epothiloneB_samples} 
\end{figure} 

\begin{figure}[!htb]
  \centering 
  \includegraphics[width=\textwidth]{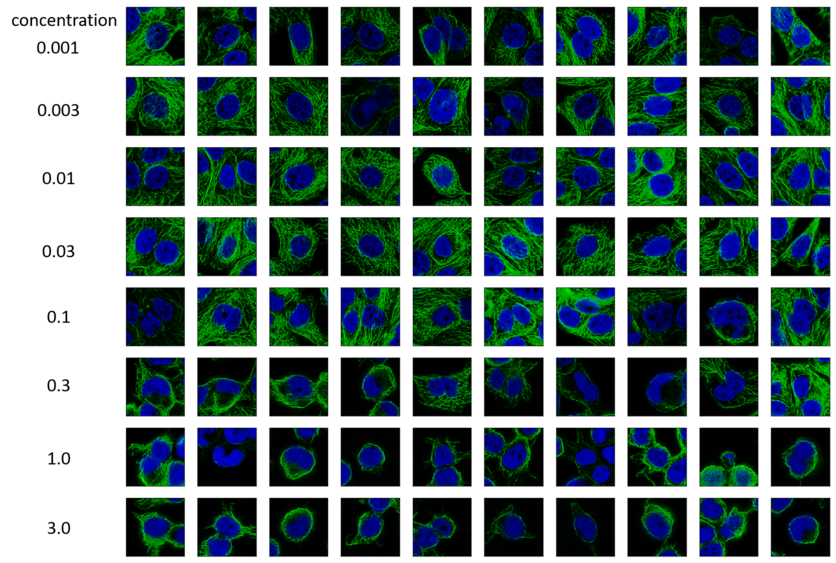} 
  \caption{Examples of reconstructed images from cells treated with Latrunculin B.}
  \label{fig:latrunculinB_samples} 
\end{figure} 

\begin{figure}[!htb]
  \centering 
  \includegraphics[width=\textwidth]{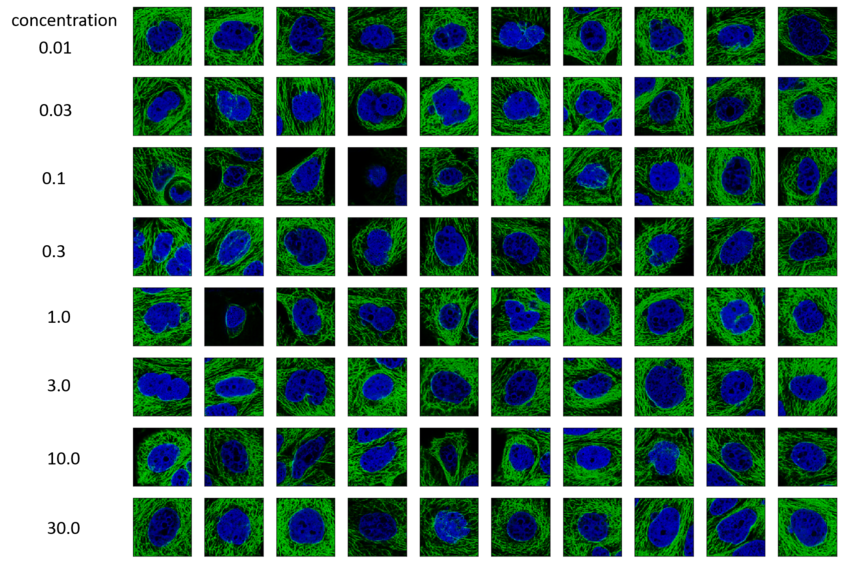} 
  \caption{Examples of reconstructed images from cells treated with AZ-A.}
  \label{fig:AZ-A_samples} 
\end{figure} 

\end{document}